\documentclass{aa1}
\usepackage{graphicx}
\usepackage{color}
\usepackage{natbib}
\usepackage{amsmath}
\usepackage{hyperref}

\begin{document}

\title{A note on identifying continuous gravitational wave emission signatures of magnetars in gamma-ray bursts}

\offprints{dushuang@pku.edu.cn;2023005@tlu.edu.cn}

\author{Shuang Du}

\institute{
    Tongling University, Tongling 244000, Anhui, China
          }

        \authorrunning{Shuang Du}
          \titlerunning{ }
\date{Received: xxx.}

\abstract{Continuous gravitational waves (GWs) of neutrons stars haven't been detected directly until now.
One possible way to indirectly identify their signatures is via the correlation between magnetars and gamma-ray bursts (GRBs),
since, under this magnetar scenario of GRBs, GW radiation can affect the evolution of GRB X-ray light curves.
Nevertheless, relevant studies lack essential details of this GRB magnetar scenario.
For instance, the authors tend to avoid answering the questions like why GRB X-ray light curves can record the information of gravitational wave emissions,
what are the reliable criteria for selecting samples that can reveal GW information, and what are the limitations of this GRB-magnetar scenario.
In this paper, we elucidate these issues in detail.
 }

\maketitle

\section{introduction}
The gravitational radiation generated by distorted neutron stars (NSs) has not been detected even in the Milky Way \citep{2014ApJ...785..119A,2017PhRvD..96f2002A,2019ApJ...875..160A,2021ApJ...913L..27A}.
This fact indicates that Galactic NSs have experienced effective spin-downs and maintain rotationally symmetric structures.
Rapid rotation and sufficient distortion are the two necessary conditions for NSs to serve as sources of gravitational waves (GWs).
Therefore, millisecond magnetars are quality sources (see, e.g., \cite{2022A&A...666A.138L} and references therein).
Some central objects of gamma-ray bursts (GRBs) are just potential candidates of millisecond magnetars.\footnote{The GW radiations associated with GRBs on other aspects can refer to
\cite{2018MNRAS.480..402D,2023ApJ...944..189H,2023MNRAS.518.5242U}.}

GRBs are bursts of gamma-rays which is identified to be originated from massive star collapses and NS binary mergers (see \cite{2018pgrb.book.....Z} for details).
Taking into account the uncertainty in the equation of state of dense matter and the abundance of energy, \cite{1986ApJ...308L..43P} proposed the idea that NSs could potentially serve as the central objects responsible for GRBs. \cite{1992Natur.357..472U} showed that if GRB central objects are distorted millisecond magnetars, there may be associations between GRBs and GW bursts. \cite{1998A&A...333L..87D} and \cite{2001ApJ...552L..35Z} introduced a popular scenario involving NSs (magnetars) to explain some GRB afterglows. Under this GRB magnetar scenario, it was suggested that the energy contributed by magnetar spin-down winds was a contributing factor to GRB X-ray plateaus (but relevant details were absent; see below). Therefore, if GW emission can effectively affect magnetar spin-downs, the evolution of GRB X-ray afterglows may record GW information.

Some authors further investigate the distortions of nascent NSs through GRB X-ray afterglows based on this idea. However, due to the lack of detailed information on how spin-down winds result in X-ray plateaus and subsequent decay segments, these works often simply assume a constant ratio between the power of the spin-down wind, $L_{\rm p}$, and the luminosity of the corresponding X-ray afterglow, $L_{\rm X}$, that is $L_{\rm X}=\eta L_{\rm p}$, leading to significant variations in the value of $\eta$ among these works ($0.01-1$; see, e.g., \citealt{2018MNRAS.480.4402L,2022MNRAS.513.1365X,2022ApJ...934..125X,2023MNRAS.522.4294Y}). \cite{2016MNRAS.458.1660L} avoid this difficulty, but only a brief constraint is obtained (see their equation (9)). Additionally, because of the lack of detailed information, all these works usually cannot present a rigorous criteria for selecting GRB samples that can reveal GW information. On the other hand, the distortion that results in GW emission can be induced by many factors, such as the distortion induced by magnetic field and fast rotation. However, authors typically consider these factors separately (for the case of magnetically induced distortion can refer to \cite{2022A&A...666A.138L,2022ApJ...934..125X}; for the case of r-mode oscillation can refer to \cite{2019ApJ...871..160L}). \cite{2022MNRAS.513.1365X} consider the magnetically induced distortion and fast-rotation-induced r-mode oscillation, but both of these two effects are still independently handled when they present an analytical formalism. The other works (e.g., \citealt{2018MNRAS.480.4402L,2023MNRAS.522.4294Y}) do not illustrate which effect causes the distortion of the NS, nor do they explain why the ellipticity of the NS remains constant.

In this paper, we aim to present the fundamental physical framework and crucial details regarding the efficiency of converting the electromagnetic energy of a magnetar into GRB X-ray emission (i.e., $\eta$). We provide a more general analytic formulation for NS spin-down which includes the contribution of the GW emission resulted by magnetically induced distortion and r-mode oscillation. Furthermore, we clarify the criteria for identifying GW emission signatures and the limitations of this approach.

\section{The basic physical frame}\label{sec2}
The luminosity of the Poynting flux emitted by a pulsar (e.g., via equivalent magnetic dipole radiation) is expected to follow the evolution
that $L_{\rm p}\propto (1+t/\tau)^{-2}$ (see, e.g., \citealt{2001ApJ...552L..35Z}), where $t$ is the time and $\tau$ is the characteristic timescale beyond which the decay of the flux accelerates.
Although the Poynting flux has the potential to be converted into X-ray emission with a luminosity $L_{\rm X}$,
the exact functional relationship between $L_{\rm p}$ and $L_{\rm X}$ is often unknown.
Consequently, when inferring the rotational evolution of the possible central NS of a GRB by analyzing the observed changes in the associated X-ray afterglow,
and subsequently investigating potential GW emission signatures,
it is necessary to assume that the evolution of X-ray afterglow luminosity precisely tracks the NS spin-down luminosity (i.e., $L_{\rm X}=\eta L_{\rm p}$ and
the value of $\eta$ must be assumed to be a constant).
Without this assumption, it would be exceedingly difficult to determine the spin-down history of the NS or comprehend the impact of GW radiation on the NS spin-down.
Besides, it is important to note that the parameter, $\eta$, in this kind of studies plays a multifaceted role.
For example, it determines the energy budget which can afford the X-ray radiation (the viability of the NS scenario of GRB X-ray afterglows),
that is
\begin{eqnarray}\label{1.1}
E_{\rm X}&=&\frac{1}{2}\eta I\Omega^{2}\nonumber\\
&\approx& 2.0\times 10^{50} {\rm erg}\left ( \frac{\eta}{0.01}  \right ) \left ( \frac{I}{10^{45}\rm g\cdot cm^{2} }  \right ) \left ( \frac{P}{1\rm ms}  \right )^{-2},
\end{eqnarray}
where $\Omega$ is the spin velocity, $I$ is the rotational inertia and $P$ is the spin period.

Undoubtedly, the parameter $\eta$ holds immense significance in obtaining meaningful results, and it should be determined based on a reasonable physical framework rather than an arbitrary assumption. Fortunately, the existence of internal plateaus offers clues for estimating the value of $\eta$ in the context of the GRB magnetar scenario, as elucidated in \cite{2020ApJ...901...75D}.
Internal plateaus are the X-ray plateaus followed by steep decay segments \citep{2010MNRAS.402..705L}.
Under the GRB magnetar scenario, when magnetars collapse into black holes there is $L_{\rm p}=0$, the contribution of the spin-down wind to the X-ray emission vanishes and
steep decay segments should appear.
Therefore, these internal plateaus are naturally corresponding to the assumption that $L_{\rm X}=\eta L_{\rm p}$.
This correspondence implies that the X-ray afterglow is produced by two independent components: the X-ray emission from the GRB jet and the X-ray emission from the spin-down wind. Therefore, the following physical schematic diagram is obtained. The energy stored in the magnetar wind is converted into X-ray emission autonomously, occurring behind the GRB jet. During the X-ray plateau phase, the primary source of energy driving the X-ray emission is the magnetar wind. When the magnetar collapses into a black hole, the contribution from the magnetar wind abruptly stops, resulting in the disappearance of the X-ray plateau. Subsequently, this transition is typically accompanied by a steep decay segment in the X-ray light curve. Following the release of a sufficient amount of energy stored in the residual magnetar wind, the contribution from the GRB jet gradually becomes apparent (corresponding to the emergence of the normal decay segment of the X-ray afterglow).
Under this scenario of GRB X-ray plateaus, the value of $\eta$ could be estimated since
the segment which is contributed by the magnetar wind may be identified, that is
\begin{eqnarray}
\eta= \frac{E_{\rm plateau}}{E_{\rm plateau}+E_{\rm remnant}},
\end{eqnarray}
where $E_{\rm plateau}$ is the total energy of the X-ray plateau, and $E_{\rm remnant}$ is the residual energy of the spin-down wind (e.g., the bulk kinetic energy).
Note that during the dissipation of Poynting flux, part of the magnetic energy in the spin-down wind will be transformed into the kinetic energy of the wind. When the accelerated wind catches up with the front jet, it may cause a bump (see Figure 1 in \citealt{2020ApJ...901...75D}). Therefore, we have $E_{\rm remnant}\sim E_{\rm bump}$ with $E_{\rm bump}$ being the energy of the bump. GRB 070110 \citep{2007ApJ...665..599T} fits this model well and can be used to estimate the value of $\eta$ ($\sim 0.9$).

\section{The analytic formalization}\label{sec3}
If one considers the spin-down of an NS which is effected by
(\textrm{i}) the magnetic dipole radiation with the power being (see Appendix \ref{ap1})
\begin{eqnarray}\label{3}
L_{\rm p}=\frac{B_{\rm eff}^{2}R^{6}\Omega^{4}}{6c^{3}},
\end{eqnarray}
where $B_{\rm eff}$ is the effective dipole magnetic field strength on the NS polar cap, $R$ is the equatorial radius, and $c$ is the speed of light;
(\textrm{ii}) the GW quadrupole radiation resulted by a magnetically induced distortion (approximates a constant during the time when the magnetic field has not yet undergone significant evolution) that \citep{PM,M08}
\begin{eqnarray}\label{1}
L_{\rm gw,\varepsilon}=\frac{32GI^{2}\epsilon_{\rm B, eff}^{2}\Omega^{6}}{5c^{5}},
\end{eqnarray}
where $G$ is the gravity constant, and $\epsilon_{\rm B, eff}$ is the effective magnetically induced ellipticity;
(\textrm{iii}) the GW radiation resulted by the leading mode of r-mode oscillation \citep{1998PhRvD..58h4020O,2016MNRAS.463..489H} that
\begin{eqnarray}\label{2}
L_{\rm gw,r}\approx \frac{96\pi}{15^{2}}\left ( \frac{4}{3} \right )^{6}
\frac{GMR^{4}\tilde{J}^{2}I}{c^{7}\tilde{I}}a^{2}\Omega^{8},
\end{eqnarray}
where $M$ is the NS mass, $\tilde{J}=\frac{1}{MR^{4}}\int_{0}^{R}\rho r^{6}dr$ with $\rho$ being the mass density,
$\tilde{I}=\frac{8\pi}{3MR^{2}}\int_{0}^{R}\rho r^{4}dr$, and $a$ is the amplitude of the oscillation,
the evolution of the spin-down should be given by
\begin{eqnarray}\label{4}
I\Omega\frac{d\Omega }{dt}=-L_{\rm p}-L_{\rm gw,\varepsilon }-L_{\rm gw,r},
\end{eqnarray}
that is
\begin{eqnarray}\label{5}
\frac{d\Omega }{dt}=\alpha \Omega^{3} +\beta \Omega^{5}+\gamma \Omega^{7},
\end{eqnarray}
where
\begin{eqnarray}
\alpha=-\frac{B_{\rm eff}^{2}R^{6}}{6c^{3}I},
\end{eqnarray}
\begin{eqnarray}
\beta=-\frac{32GI\epsilon_{\rm B, eff}^{2}}{5c^{5}},
\end{eqnarray}
\begin{eqnarray}
\gamma=-\frac{96\pi}{15^{2}}\left ( \frac{4}{3} \right )^{6}
\frac{GMR^{4}\tilde{J}^{2}}{c^{7}\tilde{I}}a^{2}.
\end{eqnarray}
The integral version of equation (\ref{5}) is
\begin{eqnarray}\label{6}
\int_{\Omega_{0}}^{\Omega_{\rm t}}\frac{d\Omega}{\alpha \Omega^{3} +\beta \Omega^{5}+\gamma \Omega^{7} }=\int_{0}^{t}dt'.
\end{eqnarray}
We note that
\begin{eqnarray}
\frac{1}{\Omega^{3}(\alpha +\beta\Omega^{2}+\gamma\Omega^{4})}=
\frac{1}{\alpha \Omega^{2}}-\frac{\beta +\gamma \Omega^{2}}{\alpha \Omega (\alpha +\beta\Omega^{2}+\gamma\Omega^{4})}\nonumber\\
=\frac{1}{\alpha \Omega^{2}}-\frac{\beta }{\alpha \Omega (\alpha +\beta\Omega^{2}+\gamma\Omega^{4})}-\frac{\gamma\Omega }{\alpha(\alpha +\beta\Omega^{2}+\gamma\Omega^{4}) }
\end{eqnarray}
and
\begin{eqnarray}
\frac{\beta }{\alpha \Omega (\alpha +\beta\Omega^{2}+\gamma\Omega^{4})}=
\frac{\beta }{\alpha^{2}\Omega}-\frac{\beta^{2}\Omega +\beta \gamma \Omega^{3} }{\alpha^{2}(\alpha +\beta\Omega^{2}+\gamma\Omega^{4})}\nonumber\\
=\frac{\beta }{\alpha^{2}\Omega}-\frac{\beta^{2}\Omega}{\alpha^{2}(\alpha +\beta\Omega^{2}+\gamma\Omega^{4})}-\frac{\beta \gamma \Omega^{3} }{\alpha^{2}(\alpha +\beta\Omega^{2}+\gamma\Omega^{4})}.
\end{eqnarray}
So, equation (\ref{6}) gives
\begin{eqnarray}\label{13}
t&=&-\frac{1}{\alpha \Omega}\bigg|_{\Omega_{0}}^{\Omega_{\rm t}}- \frac{\beta }{\alpha^{2}}\ln\Omega \bigg|_{\Omega_{0}}^{\Omega_{\rm t}}\nonumber\\
&&+\frac{\beta}{4\alpha^{2}}\ln\left | \alpha +\beta\Omega^{2}+\gamma\Omega^{4} \right | \bigg|_{\Omega_{0}}^{\Omega_{\rm t}}\nonumber\\
&&+\left(\frac{5\beta^{2}}{4\alpha^{2}}-\frac{\gamma}{\alpha} \right)\nonumber\\
&&\times\left\{\begin{matrix}
 \frac{1}{\sqrt{4 \alpha\gamma -\beta^{2}}}\arctan\frac{2\gamma\Omega^{2}+\beta }{\sqrt{4\alpha \gamma -\beta^{2}}} \bigg|_{\Omega_{0}}^{\Omega_{\rm t}} & (\beta^{2}<4\alpha \gamma) \\
\frac{1}{2\sqrt{\beta^{2}-4\alpha \gamma}}\ln\left | \frac{2\gamma\Omega^{2}+\beta -\sqrt{\beta^{2}-4\alpha \gamma} }{2\gamma\Omega^{2}+\beta+\sqrt{\beta^{2}-4\alpha \gamma}} \right|\bigg|_{\Omega_{0}}^{\Omega_{\rm t}} &(\beta^{2}>4\alpha \gamma)
\end{matrix}\right..
\end{eqnarray}

\section{Criterias and limitations}
Equation (\ref{13}) seems to be complicated.
However, the different correlations between the luminosities and angular velocity that $L_{\rm p}\propto \Omega^{4}$, $L_{\rm gw,\varepsilon}\propto \Omega^{6}$,
and $L_{\rm gw,r}\propto \Omega^{8}$ greatly alleviates the difficulties we have to face.
Since the angular velocity slows down with the time, these correlations mean that the decay of $L_{\rm gw,r}$ is fastest,
the decay of $L_{\rm gw,\varepsilon}$ is slower, and the decay of $L_{\rm p}$ is the slowest.
Hence, if the spin-down is governed by magnetic dipole radiation initially,
the GW radiation will never dominate the spin-down.
To identify the signature of GW emission in the X-ray light curve, it is vital for GW radiation to dominate the spin-down in the initial stage.
This allows the impact of GW radiation to be reflected in the X-ray light curve, ensuring its discernibility.
Under the situation of GW radiation domination, equation (\ref{4}) is reduced to
\begin{eqnarray}\label{14}
I\Omega\frac{d\Omega }{dt}\approx -L_{\rm gw,\varepsilon }-L_{\rm gw,r}.
\end{eqnarray}
In a similar vein, if the objective is to identify the GW emission signature resulted by the leading mode of r-mode oscillation,
it is imperative for the r-mode-induced GW radiation to dominate the spin-down process, that is
\begin{eqnarray}\label{15}
\left\{\begin{matrix}
I\Omega\frac{d\Omega }{dt}\approx -L_{\rm gw,r} \\
L_{\rm p}=\frac{B_{\rm eff}^{2}R^{6}\Omega^{4}}{6c^{3}}
\end{matrix}\right..
\end{eqnarray}
In this case, there is $L_{\rm X}\propto t^{-2/3}$ after the break of the X-ray plateau.
When the domination of the r-mode GW emission terminates, there are two possibilities that the following spin-down is dominated by the
magnetically-distortion-induced GW radiation (leads to $L_{\rm X}\propto t^{-1}$; see equation (\ref{16})) or dominated by the magnetic dipole radiation (leads to $L_{\rm X}\propto t^{-2}$).
Therefore, the criterion for the GW emission signature of the r-mode oscillation are that
(i) the decay index of the X-ray light curve changes as $\sim 0\rightarrow \sim -2/3\rightarrow \sim -1\rightarrow\sim -2$
(i.e., the spin-down is dominated by equation (\ref{2}), equation (\ref{1}) and equation (\ref{3}) in turn);
(ii) the decay index of the X-ray light curve changes as $\sim 0\rightarrow \sim -2/3\rightarrow\sim -2$
(i.e., the spin-down is dominated by equation (\ref{2}) and equation (\ref{3}) in turn).
Alternatively, if one wants to identify the GW emission signature resulted by the magnetically induced distortion, the initial situation
should be
\begin{eqnarray}\label{16}
\left\{\begin{matrix}
I\Omega\frac{d\Omega }{dt}\approx -L_{\rm gw,\varepsilon} \\
L_{\rm p}=\frac{B_{\rm eff}^{2}R^{6}\Omega^{4}}{6c^{3}}
\end{matrix}\right.,
\end{eqnarray}
that is $L_{\rm X}\propto t^{-1}$ follows the X-ray plateau.
In this case, the total evolution of the decay index of the spin-down wind should be $\sim 0\rightarrow \sim -1\rightarrow \sim -2$
(i.e., the spin-down is dominated by equation (\ref{1}) and equation (\ref{3}) in turn).

In order to identify a reliable candidate for shedding the information of GW radiation,
simply considering the evolution of the decay index $\sim 0\rightarrow \sim -2/3$ or $\sim 0\rightarrow \sim -1$ is not sufficient.
The reason are that (\textrm{i}) the presence of an X-ray plateau may be attributed to the GRB jet viewed slightly off-axis \citep{2020MNRAS.492.2847B},
(\textrm{ii}) decay indices within the range of $\sim (-2/3, -1)$ are commonly observed in X-ray light curves (\citealt{2007A&A...469..379E,2009MNRAS.397.1177E};
that is the decay index with the value being in $\sim (-2/3, -1)$ may be just resulted by the external shock).
The evolutionary behavior of the decay index that $\sim 0\rightarrow \sim -2/3\rightarrow\sim -2$ or $\sim 0\rightarrow \sim -1\rightarrow\sim -2$
still cannot be a definitive indicator of the GW emission signature.
This evolutionary pattern may still be clarified within the context of the jet scenario, which comprises a jet observed slightly off-axis (decay index $\sim 0$), a standard external shock phase (decay index $\in \sim (-2/3, -1)$), and an occurrence of a jet break (decay index $\sim -2$; see \citealt{2015PhR...561....1K} for a review).
In principle, the sample exhibiting an evolution pattern of $\sim 0\rightarrow \sim -2/3\rightarrow \sim -1\rightarrow\sim -2$ is a promising candidate,
as it is challenging for the jet scenario to generate such a diverse range of features.
However, from an observational perspective, identifying such a reliable sample is challenging.
To obtain a trustworthy value for the decay index through fitting the X-ray light curve, the duration of the corresponding segment must be sufficiently long.
Otherwise, due to the limited precision of the telescope, the error bars associated with the data can allow for considerable flexibility in determining the decay index.
As shown in equation (\ref{1.1}), the limited energy budget usually cannot afford sufficiently duration for every segment.
For instance, the X-ray afterglow of GRB 090426 shown in \cite{2022MNRAS.513.1365X} does not exhibit the anticipated transition of the decay index from
$\sim -1$ to$\sim -2$. This suggests that the transformation from GW radiation domination to EM radiation domination occurred at a time point, at the very least, beyond the last available data point ($> \sim 5\times 10^{5}\rm s$). Consequently, the combined energy released via GW radiation and EM radiation could surpass the energy threshold manageable by the rotational energy of the magnetar (note that $\eta\sim 0.01$ is adopted in \citealt{2022MNRAS.513.1365X}).

\begin{figure}
 \centering
 \includegraphics[width=0.4\textwidth]{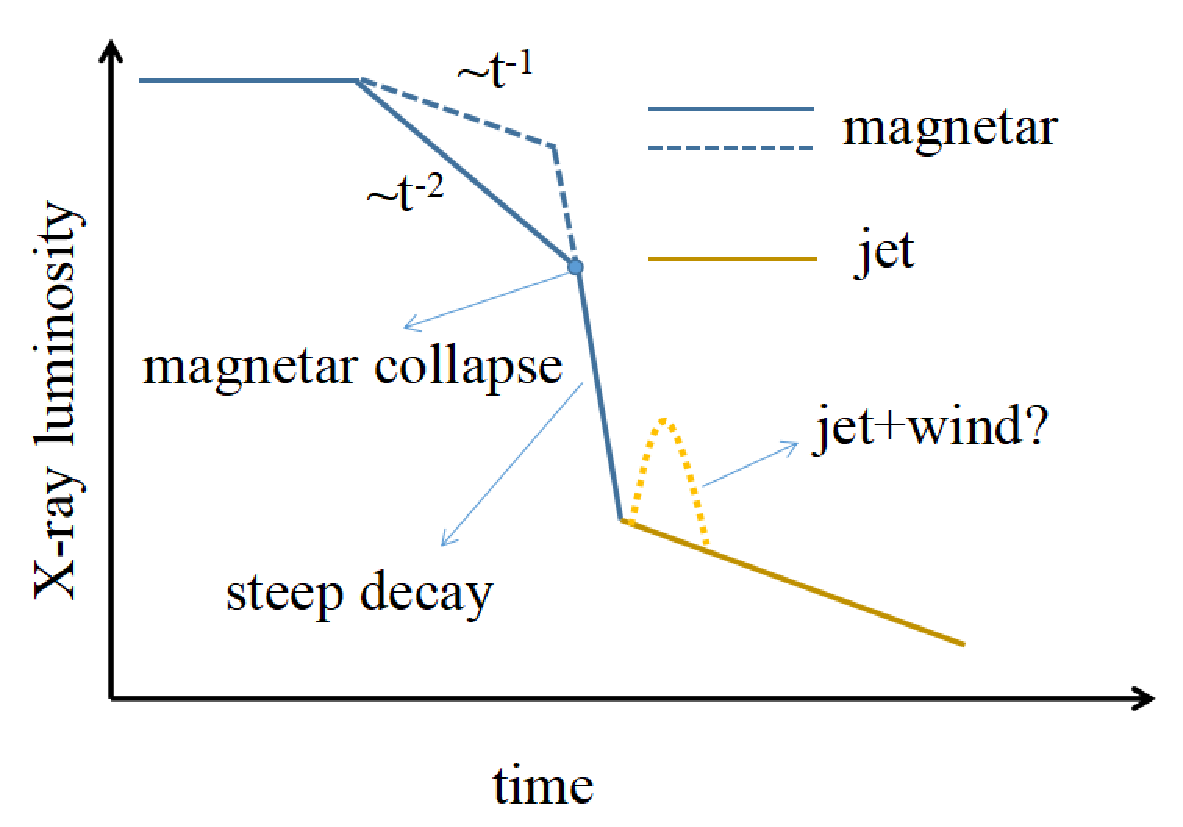}
  \caption{The schematic figure for the reliable sample of a magnetar-powered X-ray light curve.
  The dot line may be resulted by the interaction between the remnant magnetar wind and the jet \citep{2020ApJ...901...75D}.}
    \label{fig.1}
\end{figure}
Simply identifying the GW emission signature through the possible evolution of wind component is insufficient.
According to Section \ref{sec2} and above discussion,
a reliable sample of magnetar-powered X-ray light curve should clearly exhibit the wind component and jet component (see Figure \ref{fig.1}).
Although a considerable number of internal plateaus have been observed, it seems that no sample comparable to the one depicted in Figure \ref{fig.1} has been found \citep{2007A&A...469..379E,2009MNRAS.397.1177E}.
If a similar sample cannot be observed over an extended period, it suggests that the magnetar scenario of GRB X-ray plateaus should be reconsidered.

\section{Summary}
In this manuscript, we specify the basic physical framework and the general mathematical formalization which is absent in relevant works on identifying GW emission signatures in GRBs.
We present the rigorous criterion for selecting a reliable sample responsible for the GW emission signature.
We find that the current works,  which rely on varying criteria (sometimes, the authors select a sample solely based on their belief that it could potentially be a candidate),
are hard to yield a reasonable outcome (despite their similarities in conclusions; see, e.g., \citealt{2022A&A...666A.138L,2022MNRAS.513.1365X,2022ApJ...934..125X}).
However, it should be noted that studying the interiors of NSs is a challenging task,
the pursuit of such kind of investigations is beneficial.
Nevertheless, we emphasize the need for adopting a more cautious and rigorous approach in future studies.

\begin{appendix}
\section{The derivation of equation (\ref{3})}\label{ap1}
The energy flux of magnetic dipole radiation is \citep{1975clel.book.....J,LL79} (Gaussian units)
\begin{eqnarray}
\mathbf{S}=\frac{1}{4\pi c^{3}D_{\rm L}^{2}}|\ddot{\mathbf{m}}\times\mathbf{n}\times\mathbf{n}|^{2}\mathbf{n},
\end{eqnarray}
where $D_{\rm L}$ is the distance from the source to the observer, $\ddot{\mathbf{m}}$ is second derivative with respect to time of the magnetic moment $\mathbf{m}$,
and $\mathbf{n}$ is the unit vector along the line of sight.
The luminosity is given by
\begin{eqnarray}\label{sd1}
L_{\rm p}=\int \mathbf{S}d\mathbf{\Sigma}=\int SD_{\rm L}^{2}d\Theta,
\end{eqnarray}
where $d\mathbf{\Sigma}$ is the surface element, and $d\Theta$ is the element of solid angle.
According to the geometry shown in Figure 2 in \cite{2019ApJ...886...87D}, equation (\ref{sd1}) can be integrated directly.
Another easier method is by means of the properties of tensors.
Note that, there is
\begin{eqnarray}\label{a16}
&&|\ddot{\mathbf{m}}\times\mathbf{n\times\mathbf{n}}|^{2}\nonumber\\
&=&\ddot{m}_{x}^{2} n_{x}^{2} n_{y}^{2} + \ddot{m}_{x}^{2} n_{x}^{2} n_{z}^{2} + \ddot{m}_{x}^{2}n_{y}^{4} + 2\ddot{m}_{x}^{2}n_{y}^{2}n_{z}^{2} + \ddot{m}_{x}^{2}n_{z}^{4}\nonumber\\
& -& 2\ddot{m}_{x}\ddot{m}_{y}n_{x}^{3}n_{y} - 2\ddot{m}_{x}\ddot{m}_{y}n_{x}n_{y}^{3} - 2\ddot{m}_{x}\ddot{m}_{y}n_{x}n_{y}n_{z}^{2}\nonumber\\
 &-& 2\ddot{m}_{x}\ddot{m}_{z}n_{x}n_{y}^{2}n_{z} - 2\ddot{m}_{x}\ddot{m}_{z}n_{x}n_{z}^{3} + \ddot{m}_{y}^{2}n_{x}^{4} + \ddot{m}_{y}^{2}n_{x}^{2}n_{y}^{2}\nonumber\\
& +& \ddot{m}_{y}^{2}n_{y}^{2}n_{z}^{2} + \ddot{m}_{y}^{2}n_{z}^{4} - 2\ddot{m}_{y}\ddot{m}_{z}n_{x}^{2}n_{y}n_{z} - 2\ddot{m}_{y}\ddot{m}_{z}n_{y}^{3}n_{z}\nonumber\\
 &+& \ddot{m}_{z}^{2}n_{x}^{4} + 2\ddot{m}_{z}^{2}n_{x}^{2}n_{y}^{2} + \ddot{m}_{z}^{2}n_{x}^{2}n_{z}^{2} + \ddot{m}_{z}^{2}n_{y}^{4} + \ddot{m}_{z}^{2}n_{y}^{2}n_{z}^{2}\nonumber\\
 & -& 2\ddot{m}_{x}\ddot{m}_{z}n_{x}^{3}n_{z}+ 2\ddot{m}_{y}^{2}n_{x}^{2}n_{z}^{2} -2\ddot{m}_{y}\ddot{m}_{z}n_{y}n_{z}^{3},
\end{eqnarray}
where ${m}_{ i}\;(i=x,y,z)$ is the component of ${\mathbf{m}}$, and $n_{ i}$ is the component of $\mathbf{n}$.
Via Cauchy representation theorem of tensors, one can construct, for example,
\begin{eqnarray}
\overline{n_{ i}n_{ j}n_{ l}n_{ k}}&\equiv& \frac{1}{4\pi}\int_{0}^{4\pi} n_{ i}n_{ j}n_{ l}n_{ k}d\Theta\nonumber\\
&=&\xi_{1}\delta_{ i j}\delta_{ l k}+\xi_{2}\delta_{ i l}\delta_{ j k}+\xi_{3}\delta_{ i k}\delta_{ j l}
\end{eqnarray}
since $\overline{n_{i}n_{j}n_{ l}n_{k}}$ is invariable under the rotation of the coordinate system (briefly, one can consider that the flux is invariable under the rotation of the coordinate system),
where $\xi_{1},\xi_{2},\xi_{3}$ are constants.
On the other hand, $\overline{n_{ i}n_{ j}n_{ l}n_{ k}}$ is symmetric, so that there should be
\begin{eqnarray}
\overline{n_{ i}n_{ j}n_{ l}n_{ k}}
=\frac{1}{15}(\delta_{ i j}\delta_{ l k}+\delta_{ i l}\delta_{ j k}+\delta_{ i k}\delta_{ j l}),
\end{eqnarray}
where the constant, $\frac{1}{15}$, comes from that the right-hand side of this equation should equal to $1$ (i.e., $\frac{1}{4\pi}\int_{0}^{4\pi} d\Theta$) after the contractions of $i, j$ and $l, k$.\footnote{Readers may refer to Exercise 45 in Section 9.7 of \cite{SB} for a similar consideration.}
Regardless of the derivation of the equation above, there is a general mathematical formula that\footnote{I cannot recall which book I encountered this formula in.}
\begin{eqnarray}\label{a19}
&&\frac{1}{4\pi}\int n_{k_{1}}\cdots n_{k_{m}}d\Theta \nonumber\\
&=&\begin{cases}
 & \frac{m(\delta _{k_{1}k_{2}}\cdots\delta _{k_{m-1}k_{m}}+\mathrm{permutations})}{(m+1)!}\text{ if } m= \mathrm{even} \\
 & 0 \;\;\;\;\;\;\;\;\;\;\;\;\;\;\;\;\;\;\;\;\;\;\;\;\;\;\;\;\;\;\;\;\;\;\;\;\; \;\;\;\;\;\;\;\text{ if } m= \mathrm{odd}
\end{cases}.
\end{eqnarray}
According to equation (\ref{a19}),
the nonzero terms of right-hand side of equation (\ref{a16}) are given by
\begin{eqnarray}
\overline{\ddot{m}_{i}^{2}n_{i}^{2}n_{j}^{2}}=\frac{1}{15}\ddot{m}_{i}^{2},
\end{eqnarray}
\begin{eqnarray}
\overline{\ddot{m}_{i}^{2}n_{j}^{4}}=\frac{3}{15}\ddot{m}_{i}^{2},
\end{eqnarray}
and
\begin{eqnarray}
\overline{\ddot{m}_{i}^{2}n_{j}^{2}n_{k}^{2}}=\frac{1}{15}\ddot{m}_{i}^{2}.
\end{eqnarray}
Therefore, we get
\begin{eqnarray}\label{a23}
L_{\rm p}=\frac{1}{ c^{3}}\sum_{i\neq j\neq k}(\overline{\ddot{m}_{i}^{2}n_{i}^{2}n_{j}^{2}}+\overline{\ddot{m}_{i}^{2}n_{j}^{4}}+\overline{\ddot{m}_{i}^{2}n_{j}^{2}n_{k}^{2}})=\frac{2\ddot{m}^{2}}{3c^{3}}.
\end{eqnarray}
For the magnetic dipole field (see, e.g., \citealt{1975clel.book.....J})
\begin{eqnarray}\label{coj2}
\mathbf{B}_{\rm p}=\frac{3\mathbf{n}(\mathbf{n}\cdot \mathbf{m})-\mathbf{m}}{R^{3}},
\end{eqnarray}
we have (see Figure 2 in \citealt{2019ApJ...886...87D})
\begin{eqnarray}\label{a24}
\mathbf{m}&=&\frac{B_{\rm p}R^{3}}{2}\nonumber\\
&&\times (\mathbf{e}_{\parallel}\cos\alpha +\mathbf{e}_{1}\sin\alpha\cos\Omega t+\mathbf{e}_{2}\sin\alpha\sin\Omega t),
\end{eqnarray}
where $B_{\rm p}=B_{\rm eff}/\sin\alpha$ (in this Appendix, $\alpha$ is an angle).
Substituting equation (\ref{a24}) into equation (\ref{a23}), we obtain equation (\ref{3}).

\section{A comment on Xie et al. (2022a)}\label{ap2}
When I read \cite{2022MNRAS.513.1365X}, I find some errors and ambiguous contents which may mislead readers (especially for beginners).
I believe it is worthwhile to elaborate on these errors.

(\textrm{I}) Xie et al. mentioned that the energy injection by the internal dissipation of the magnetar wind powered the corresponding X-ray plateau,
and cited others' works, such as \cite{2011MNRAS.413.2031M}.
However, the concept ``energy injection" in \cite{2022MNRAS.513.1365X} is ambiguous, at least, \cite{2011MNRAS.413.2031M} did not mention this concept.
The concept ``energy injection" is usually based on the work of \cite{1998A&A...333L..87D},
where this concept means that the GRB fireball obtains/absorbs energy from the pulsar wind.
Then, this concept is incompatible with GRB X-ray plateaus, especially internal plateaus emphasized in \cite{2022MNRAS.513.1365X} (see \citealt{2020ApJ...901...75D} for details).

(\textrm{II}) Xie et al. took $\eta\sim 0.01$ for all samples.
They stated that the practice was based on the result of \cite{2019ApJ...878...62X}.
However, regardless of whether the physical basis of \cite{2019ApJ...878...62X} is correct,\footnote{A strict method to constrain the value of $\eta$ can refer to \cite{2019ApJ...886...87D}.}
or even if it is, this parameter cannot still be a constant for all samples according to their conclusion.
The result of \cite{2019ApJ...878...62X} states that ``this efficiency strongly depends on the injected luminosity''. Therefore,
$\eta$, should not be a constant for different samples since periods and magnetic fields of different NSs can be much different.
Besides, as shown in \cite{2011MNRAS.413.2031M} (the paper also cited in Xie et al. (2022a)), the magnetization factor changes with time in a large value domain.
So, as emphasized in \cite{2022MNRAS.513.1365X} that the efficiency depends largely on the initial magnetization factor of magnetar wind,
$\eta$ can not be a constant.

(\textrm{III}) As discussed in Section \ref{sec3}, GRB 090426 discussed in \cite{2022MNRAS.513.1365X} cannot be a candidate for the magnetar central engine.
GRB 150424A cannot also be considered as the case that the spin-down evolution is dominated by GW radiation initially.
According to the bottom right panel of Figure 2 shown in \cite{2022MNRAS.513.1365X}, the observation does not show a plateau segment.
They fabricated a plateau segment that didn't exist.
Besides, in their case (ii), Xie et al. mentioned that the light curve was fitted with a broken power-law function.
However, they did not show the function (One can refer to equation (14) in \cite{2010ApJ...715..477Y} and equation (20) in \cite{2022A&A...666A.138L}).
The value of $\gamma_{\rm r}$ (i.e., $\alpha$) was also not stated in their fitting. In \cite{2022MNRAS.513.1365X}, $\alpha$ has three means that amplitude of r-mode, decay index of X-ray light curve and $\alpha$ effect in $\alpha-\Omega$ dynamo mechanism (this is confusing).

(\textrm{IV}) Due to the limited sensitivity of \emph{Swift} XRT, usually, it is hard to get such precise values for X-ray luminosities as shown in Table 1 of \cite{2022MNRAS.513.1365X}.
The symbol, $\sim$, should be incompatible with the number, such as, $2.55e45$.

(\textrm{V}) In the last paragraph of Section 3 of \cite{2022MNRAS.513.1365X}, the authors stated that ``It is
worth noting that no r-mode radiation candidates have been found in the observational data of SGRBs''.
However, according to the criterion shown in Section 2 of \cite{2022MNRAS.513.1365X}, GRB 150424A should just be the candidate of r-mode radiation
but not the candidate of the GW quadrupole radiation induced by constant distortion
since $|-0.81+\frac{2}{3}|$ is smaller than $|-0.81+1|$.\footnote{Why r-mode
instability is unlikely to play a non-trivial role in the evolution of magnetar can simply refer to the first paragraph of Section 2 of \cite{2022A&A...666A.138L}.}
Besides, according to their criterion, the sample shown in \cite{2019ApJ...871..160L} should be also the candidate of r-mode radiation.

(\textrm{VI}) The authors of \cite{2022MNRAS.513.1365X} made a mistake when they cited equations (6) and (7) from \cite{2016MNRAS.463..489H}.
As shown in \cite{2022MNRAS.513.1365X}, they said `` we take $\Gamma=m=2$ harmonic mode with $R=9.99\rm  km$ to $\cdots$" (below their equation 8).
In \cite{2016MNRAS.463..489H}, $\Gamma=2$ is the parameter related to the equation of state of NSs (i.e., $2$ in $p=k \rho^{2}$ ; beyond equation 2.12 of \cite{1998PhRvD..58h4020O}), but not a harmonic mode.
On the other hand, $m=2$ in \cite{1998PhRvD..58h4020O} is the ``spin quantum number'' ($l=2$ is the ``quantum number of orbital angular momentum'' ).
\textbf{Although $\Gamma$ and $m$  have the same value, they have different physical meanings, that is ``$\Gamma= m = 2$'' is forbidden.}

(\textrm{VII}) In \cite{2022MNRAS.513.1365X}, the result that $\varepsilon<1.58\times 10^{-3}(B/10^{15}\rm G)(P/1\rm ms)$ (the important content shown in their abstract; see also equation (10) in \cite{2016MNRAS.463..489H})
should be
\begin{eqnarray}
\varepsilon<1.58\times 10^{-3}\times \left ( \frac{R}{9.99\rm km}  \right )^{3} \left ( \frac{I}{1.91\times 10^{45}\rm g cm^{2}}  \right )^{-1}\nonumber\\
\times \left ( \frac{B}{10^{15}\rm G}  \right )\left ( \frac{P_{0}}{1\rm ms}  \right )\left ( \frac{\sin\theta}{1}  \right ) ^{2}.
\end{eqnarray}
according to their equation (11) and the content below their equation (4).
According to this constraint, $\varepsilon$ should be related to $P_{0}$.
Since the values of different initial periods are different, this means $\varepsilon$ is related to different values of periods,
that is $\varepsilon$ has an additional evolution versus the time since periods decay with time.
Note that, according to equation (9) in \cite{2022MNRAS.513.1365X}, the authors had acquiesced in that $\varepsilon$ is a constant.

Although the three papers that \cite{2022A&A...666A.138L,2022MNRAS.513.1365X,2022ApJ...934..125X} have similar conclusions (and even similar
procedures), I only recommend \cite{2022A&A...666A.138L} here since this paper is much more clearer (even if include \cite{2018MNRAS.480.4402L,2023MNRAS.522.4294Y}).

\end{appendix}


\section*{Acknowledgments}
I thank the organizers of the annual meeting ``gamma-ray bursts and related frontier physics'' held in Kunming, Yunnan, China, in January 2018, which allows me to
share the method that how to estimate ellipticities of NSs through GRB observations (the relation between GRB X-ray afterglows and GW radiations)
with other\textbf{s} (see also the acknowledgement in \cite{2019ApJ...886...87D}).
I also thank Peking University provide comfortable conditions that allow me to solve the problem on $\eta$ in \cite{2020ApJ...901...75D},
so, we can realize the method shown by equations (9-14) of \cite{2019ApJ...886...87D} in \cite{2022A&A...666A.138L}.
I thank myself for my interest in physics so that I can be happy and contented.
This work is supported by a research start-up fund of Tongling University.


\begin{thebibliography}{}
\bibitem[Aasi et al.(2014)]{2014ApJ...785..119A} Aasi, J., Abadie, J., Abbott, B.~P., et al. 2014, ApJ, 785, 119
\bibitem[Abbott et al.(2017)]{2017PhRvD..96f2002A} Abbott, B.~P., Abbott, R., Abbott, T.~D., et al. 2017, PhRvD, 96, 062002
\bibitem[Abbott et al.(2019)]{2019ApJ...875..160A} Abbott, B.~P., Abbott, R., Abbott, T.~D., et al.\ 2019, ApJ, 875, 160
\bibitem[Abbott et al.(2021)]{2021ApJ...913L..27A} Abbott, R., Abbott, T.~D., Abraham, S., et al.\ 2021, ApJL, 913, L27
\bibitem[Beniamini et al.(2020)]{2020MNRAS.492.2847B} Beniamini, P., Duque, R., Daigne, F., et al.\ 2020, MNRAS, 492, 2847
\bibitem[Dai \& Lu(1998)]{1998A&A...333L..87D} Dai, Z.~G. \& Lu, T.\ 1998, A\&A, 333, L87
\bibitem[Du et al.(2018)]{2018MNRAS.480..402D} Du, S., Li, X.-D., Hu, Y.-M., et al.\ 2018, MNRAS, 480, 402
\bibitem[Du et al.(2019)]{Du2019}Du, S., Peng, F. K., Long, G. B., Li, M., 2019, MNRAS, 482, 2973
\bibitem[Du et al.(2019)]{2019ApJ...886...87D} Du, S., Zhou, E., \& Xu, R.\ 2019, ApJ, 886, 87
\bibitem[Du(2020)]{2020ApJ...901...75D} Du, S.\ 2020, ApJ, 901, 75
\bibitem[Evans et al.(2007)]{2007A&A...469..379E} Evans, P.~A., Beardmore A.~P., Page K.~L., et al.\ 2007, A\&A, 469, 379
\bibitem[Evans et al.(2009)]{2009MNRAS.397.1177E} Evans, P.~A., Beardmore A.~P., Page K.~L., et al.\ 2009, MNRAS, 397, 1177
\bibitem[Huang et al.(2023)]{2023ApJ...944..189H} Huang, B.-Q., Liu, T., Xue, L., et al.\ 2023, ApJ, 944, 189
\bibitem[Ho(2016)]{2016MNRAS.463..489H} Ho, W.~C.~G.\ 2016,MNRAS, 463, 489
\bibitem[Jackson(1975)]{1975clel.book.....J} Jackson, J.~D.\ 1975, 92/12/31, New York: Wiley, 1975, 2nd ed.
\bibitem[\protect\citeauthoryear{Kumar \& Zhang}{2015}]{2015PhR...561....1K} Kumar P., Zhang B., 2015, PhR, 561, 1
\bibitem[Landau \& Lifshitz(1979)]{LL79} Landau L. D., \& Lifshitz E. M., 1979, Course of Theoretical Physics, Vol. 2: The Classical Theory of Fields, Pergamon Press, Oxford
\bibitem[\protect\citeauthoryear{Lasky \& Glampedakis}{2016}]{2016MNRAS.458.1660L} Lasky, P.~D., Glampedakis K., 2016, MNRAS, 458, 1660
\bibitem[Lin \& Lu(2019)]{2019ApJ...871..160L} Lin, J. \& Lu, R.-J.\ 2019, ApJ, 871, 160
\bibitem[Lin et al.(2022)]{2022A&A...666A.138L} Lin, T., Du, S., Wang, W., et al.\ 2022, A\&A, 666, A138
\bibitem[Lyons et al.(2010)]{2010MNRAS.402..705L} Lyons, N., O'Brien, P.~T., Zhang, B., et al.\ 2010, MNRAS, 402, 705
\bibitem[L{\"u} et al.(2018)]{2018MNRAS.480.4402L} L{\"u}, H.-J., Zou, L., Lan, L., et al.\ 2018, MNRAS, 480, 4402
\bibitem[Maggiore(2008)]{M08} Maggiore M., 2008, Gravitational Waves, Vol. 1: Theory and Experiments, Oxford University Press
\bibitem[Metzger et al.(2011)]{2011MNRAS.413.2031M} Metzger, B.~D., Giannios, D., Thompson, T.~A., et al.\ 2011, MNRAS, 413, 2031
\bibitem[Owen et al.(1998)]{1998PhRvD..58h4020O} Owen, B.~J., Lindblom, L., Cutler, C., et al.\ 1998, PhRvD, 58, 084020
\bibitem[\protect\citeauthoryear{Paczynski}{1986}]{1986ApJ...308L..43P} Paczynski, B. 1986, ApJL, 308, L43
\bibitem[Peters \& Mathews(1963)]{PM} Peters, P.C., Mathews, J. 1963, Physical Review, 131, 435
\bibitem[Schutz(2009)]{SB}Schutz, B. 2009, A First Course in General Relativity. Cambridge University Press
\bibitem[Troja et al.(2007)]{2007ApJ...665..599T} Troja, E., Cusumano, G., O'Brien, P.~T., et al.\ 2007, ApJ, 665, 599
\bibitem[Urrutia et al.(2023)]{2023MNRAS.518.5242U} Urrutia, G., De Colle, F., Moreno, C., et al.\ 2023, MNRAS, 518, 5242
\bibitem[Usov(1992)]{1992Natur.357..472U} Usov, V.~V.\ 1992, Nature, 357, 472
\bibitem[Xiao \& Dai(2019)]{2019ApJ...878...62X} Xiao, D. \& Dai, Z.-G.\ 2019, ApJ, 878, 62
\bibitem[Xie et al.(2022a)]{2022MNRAS.513.1365X} Xie, L., Wei, D.-M., Wang, Y., et al.\ 2022a, MNRAS, 513, 1365
\bibitem[Xie et al.(2022b)]{2022ApJ...934..125X} Xie, L., Wei, D.-M., Wang, Y., et al.\ 2022b, ApJ, 934, 125
\bibitem[Yuan et al.(2023)]{2023MNRAS.522.4294Y} Yuan, Y., Fan, X.-L., \& L{\"u}, H.-J.\ 2023, MNRAS, 522, 4294
\bibitem[Yu et al.(2010)]{2010ApJ...715..477Y} Yu, Y.-W., Cheng, K.~S., \& Cao, X.-F.\ 2010, ApJ, 715, 477
\bibitem[Zhang \& M{\'e}sz{\'a}ros(2001)]{2001ApJ...552L..35Z} Zhang, B. \& Meszaros, P.\ 2001, ApJL, 552, L35
\bibitem[Zhang(2018)]{2018pgrb.book.....Z} Zhang, B.\ 2018, \emph{The Physics of Gamma-Ray Bursts}. ISBN: 978-1-139-22653-0. Cambridge Univeristy Press
\end{thebibliography}
\end{document}